# Desperately seeking the impact of learning analytics in education at scale: Marrying data analysis with teaching and learning

Olga Viberg[1] & Åke Grönlund[2]

[1] KTH Royal Institute of Technology, Department of Media Technology and Interaction Design, Sweden. E-mail: oviberg@kth.se

[2] Örebro University, Örebro University School of Business, Sweden. E-mail: ake.gronlund@oru.se

**Authors**

**Olga Viberg** is Associate Professor in Media Technology with specialization in Technology-Enhanced Learning, Department of Media Technology and Interaction Design, KTH Royal Institute of Technology, Stockholm, Sweden. Her fields of expertise include technology-enhanced learning, self-regulated learning, learning analytics, mobile learning, learning design, and design science research in education. Her work has been published in many high ranked referred journals and conferences. She is an active member of the Society for Learning Analytics Research and the European Association of Technology-Enhanced Learning.

**Åke Grönlund** is Professor in Informatics, Örebro University School of Business, Department of Informatics, Örebro, Sweden. Åke's research concerns the use of digital in various human activities. The common denominator involved in all projects is to understand how people arrange their work, their organizations, and other activities pertaining to private life, such as socializing on the web, and how technology can be used to make improvements. In particular, focus is on the fields of education and e-government.

# Introduction

Learning analytics (LA) - "the measurement, collection, analysis and reporting of data about learners and their contexts, for purposes of understanding and optimizing learning and the environments in which it occurs" (Long & Siemens, 2011, p. 34) – has the potential to impact student learning, at scale (Knight, Gibson, & Shibani, 2020) by offering critical insights into the processes of online and face-to-face learning, as well as supporting learning activities via analytics tools (Ochoa, Knight, & Wise, 2020). However, the potential is far from realized due to several interrelated issues, including "lack of uptake and adoption, generalizability, and the relevance of some of our [researchers'] work to actual practice and the power dynamics of data-informed approaches" (Ochoa et al., 2020, p. 2). Moreover, scholars highlight that to be able to improve the quality of teaching and learning at scale, we also need to consider stakeholders' data literacy – i.e., the ability to understand, find, collect, interpret, visualize, and support arguments using quantitative and qualitative data (Deahl, 2014) – and leadership (Schildkamp, 2019; Henderson & Corry, 2020).

There is still very little existing empirical evidence of LA research having improved learning, learner support and teaching at scale. Viberg et al. (2018), for example, in their systematic literature review of LA found that only nine percent of 252 reviewed studies, conducted in higher education, showed that LA improved students' learning outcomes. Also, despite the increasing availability of student data in K-12 settings worldwide – accelerated by the forced move to online education during the pandemics – the majority of LA analyses have hitherto been conducted in the context of higher education, often at a small scale within a context of one course, or in the setting of massive online open courses (MOOCs; e.g., Yu, 2021).

There are many reasons behind the slow adoption of LA and data-driven decision making processes in educational settings, especially in K-12 education, including challenges related to data interoperability (Dodero et al., 2017; Samuelsen, Chen, & Wasson, 2019), ethics and privacy concerns (Beerwinkle, 2020; Livingstone, 2020; Viberg, Andersson et al., 2021), development of stakeholders' data literacy (Ifenthaler et al., 2020) as well as feedback literacy skills (Jivet, 2021), and a general lack of participatory approaches that take into account the needs and preferences of the students and teachers – even less actually engage them directly – in the LA design process (Buckingham Shum, Ferguson, & Martinez-Maldonado, 2019; Jivet, 2021). We should not forget that LA is about *supporting* learning, not just reporting it (Gasevic, Dawson, & Simiens, 2015). Yet, the majority of LA analyses have so far been driven by the availability of learner data, often in the context of MOOCs – rather than by specific needs of teachers and students.

In this chapter, we argue that in order to make LA attractive to educational professionals and students there is a need to further develop a *human-centered learning analytics* approach. This approach posits that the design process of effective LA must extend beyond sound technical and pedagogical principles; it needs to carefully consider a range of contextual and human factors, including *why* and *how* they will be used (Buckingham Shum et al., 2020), as well as by *whom* and in what *context*. Moreover, it must be designed for the benefit of the users rather than imposed upon them by designers or researchers. In particular, we focus our discussion on the need to understand, engage and support *teachers* – the key stakeholders who guide and facilitate learning activities in everyday education practices and who are are responsible for the design and real-time management of students' learning processes (van Leeuwen et al., 2017) – as *enablers* and *co-designers* of LA. This is important, since if LA is to be able to help them through the provision of improved teacher support, we need to: 1) carefully analyse teachers' needs before implementing LA, 2) understand what data are needed and to meet those needs, 3) investigate what type of LA support mechanisms they would like to be assisted by, 4) understand how teachers would like such LA tools (e.g., teacher-facing learning dashboards) to be designed and used. Also, there is a need to understand what knowledge and skills (e.g., data and feedback literacy) they may need to develop so as to be able to enable everyday LA practices and use the designed tools effectively. These steps need to be taken in order to move towards the ultimate goal of achieving the intended changes in students' learning behaviors that would lead to improved learning outcomes.

# Critical aspects of LA in a human-centered perspective

In this section we review the literature organized by aspects that are critical in a human-centered responsible perspective of LA. The aspects include:

- *Focus on teachers' needs and goals*. This should be the starting point for LA development as this is where tools are to be applied. Unless teachers can make sense of data and take action on it, it is not meaningful.
- *Teachers' data literacy skills*. Data presentation and analysis can reveal new patterns and inspire teaching and support learning in new ways.
- *Data*. Data sources and data analysis need to become more diverse and develop beyond summative analysis of performance towards formative guidance. In doing so also students and learning contexts must be represented.

**Focus on teachers' needs and goals**

In order to take advantage of the affordances of LA to assist teachers and make a difference in their everyday teaching practice, we need to enable them to conduct meaningful – i.e., such that can make a positive difference – LA activities in their everyday practices.

There is also an aspect of responsibility involved; protection of students' privacy, is seen as both a moral and legal obligation. To achieve this, there is a need to increase our understanding of what problems they encounter in their daily teaching practice and how LA can provide assistance in addressing those problems. One of the key issues in this regard relates to the fact that data use is mainly used for *accountability* purposes, i.e., data collection and use focuses on achievement and not learning (Mandinach & Schildkamp, *in press)*. For example, test results are most often used summatively to measure performance but in order to improve learning it is more productive to use them formatively to guide teaching in a way that continuously improves conditions for student learning. Understanding *why* students have not learned something is then more interesting than the performance measure itself, but also more challenging to derive from data. In daily practice, teachers regularly assess students' work formatively – both through informal everyday observations and through continuous assessment of different individual and collaborative learning activities – in order to help them focus their thinking or improve their work processes. In formal systems, to the contrary, test results are registered for the purpose of evaluation and comparison – often of both students and teachers. While measuring student performance is necessary, it may take up too much attention at the detriment of learning. If there "is too much accountability pressure, this often leads to misuse of data, and even to abuse" (Mandinach & Schildkamp, *in press,* p. 4). It may also lead to mistrust among teachers towards LA systems as they are seen as tools for control rather than for assistance.

To provide assistance that would improve students' learning by offering adequate teacher support mechanisms, some LA researchers argue for the need of a "subversive" LA that aims to grapple with the ramifications of LA research efforts, and critically engage with the ways in which power, race, gender and class influence and are influenced by LA work (Wise, Sarmiento, & Boothe, 2021). This becomes especially critical in the context of the recent turn to online learning worldwide, which has contributed to the considerably increased *quantity* and also *kinds* of data being generated about students and by students, at different levels of education. This in turn, requires LA researchers and practitioners to address questions about bias, equity, surveillance, ownership (of data), control and agency in LA use (Wise et al., 2021).

When developing LA tools, designers need to take a critical view by integrating ways to look at the shortcomings of data in their thinking and by incorporating stakeholder privacy protection mechanisms in the tools (Klein, 2021). Such experiences and knowledge can be adapted from relevant efforts in other research areas. This includes *human-computer interaction*, which focuses on the user- and human-centered design processes aiming to effectively meet user/stakeholder needs, and *information systems research*, in which LA scholars can draw on the affordances of relevant theoretical contributions which may help underpin the conceptual groundings of LA technological artifacts.

For example, when developing LA tools – aiming to assist teachers in their work – an *ensemble view* of technology (Orlikowski & Iacono, 2001) can be adapted. This perspective focuses on the transformational nature of technology: Technology brings changes not only in terms of *how* we do things but also in terms of *what is doable and desirable* in teaching and learning practice; that is, in teachers' perceptions. This view undertakes the idea of sociotechnical construction, suggesting that new data-driven, in the context of this chapter, teaching practices and methods are co-constructed in a sociotechnical system rather than purely engineered by developers. Technology develops during use, and based on how it is used. This implies that there is a need to pay attention to the contexts of teaching and learning and to student behavior, preferences and individual characteristics. These are complex phenomena which are difficult to derive from data. Teachers regularly interpret both students and situations, but they do that not just based on structured data but also on understanding, which comes from their teaching experience, informal observations, and familiarity with students and learning situations. A teacher knows what a student *is*, the computer knows only (partially) what s/he says and does.

Data existing in digital format are typically quite simple while teachers' interpretations are based on both quantitative and qualitative data. While LA data typically informs about student scores on a test, teachers can understand *why* a student scored good or poor. Of course, LA systems could add more data. Student behavior data could be expanded to include, e.g., interaction tracing by combining data originated from different online learning platforms and also qualitative data, e.g., about students' level of motivation. But more data alone does not create understanding. There also needs to be a way to interpret data and what it tells us about student learning; that is to translate awareness into action.. Artificial intelligence (AI) can help understanding student development in for example, mathematics and literacy, but the reasons for this development, or lack thereof cannot be understood.

A major threshold in AI development today is precisely that; it can create texts that look like human-written text but as long as it lacks the understanding of what the text *means* there is a limit to its further development (Hao, 2020). In learning, teachers interacting with students are the only ones who can develop understanding and hence understand how to use data and automated analysis of data for improved learning.

Hence, a coherent understanding requires careful analysis of learning contexts and actors so as to be able to address critical questions about (algorithmic) bias, equity, surveillance, ownership (of data), control and agency in LA design and use (Wise et al., 2021). Contextual understanding may also require taking into account qualitative data, yet largely unexplored by the LA community, for example cultural differences (e.g., in terms of power distance) that may influence the adoption and the effectiveness of LA interventions (see e.g., Davis et al., 2017; Kizilcec & Cohen, 2017).

To sum up, it is critical that data use or LA activities start with a certain improvement goal and not a sole emphasis on accountability or on the available data. Taking the teacher perspective, in today's society such goals should focus not only on the improvement of students' subject knowledge, but also on development of their critical 21st century skills, including collaborative and self-regulated learning skills that are directly associated with academic performance, especially in online learning settings (e.g., Viberg, Khalil, & Baars, 2020). Moreover, such goals may be directed towards the development and improvements of students' data-, feedback- and digital literacy skills that are crucial for their successful navigation and study success in online learning settings (see e.g., Ifenthaler et al., 2020; Jivet, 2021).

**Teachers' data literacy skills**

While understanding teachers, students, and learning processes is important for LA tools designers, the increased use of data also puts demands on teachers' data literacy skills (Henderson & Corry, 2020). They need to be able to interpret data from those tools and combine it with their pedagogical knowledge so as to make it actionable in educational practices (Gummer & Mandinach, 2015). Data literacy skills include the ability to understand what data are needed to address a specific problem, collect these data, make sense of (student) data representations and feedback provided through LA tools, and based on this sense making provide improved student assistance. Research shows that educators frequently struggle with the use of data, including setting up clear goals for improvement, collecting data and making sense of them (Mandinach & Schildkamp, *in press*). In general, scholars argue that educators must have some level of data literacy, which refers to "the ability to transform information into actionable knowledge and practices by collecting, analyzing, and interpreting all types if data (assessment, school climate, behavioral, snapshot, longitudinal, moment-to-moment, etc.) to help determine instructional steps. It combines an understanding of data with standards, disciplinary knowledge and practices, curricular knowledge, pedagogical content knowledge, and an understanding of how [students] learn" (Mandinach & Gummer, 2016, p.14).

**Data**

Data-based decision making has emerged and evolved in education for almost two decades. Understanding how to make use of data in educational settings and for educational purposes is "a complex and interpretative process, in which goals have to be set, data have to be identified, collected, analyzed, and interpreted, and used to improve teaching and learning" (Mandinach & Schildkamp, in press, p.1). It is critical to acknowledge that this process does not start with data but with goals to be achieved.

To address teachers' needs, there is also a necessity to understand what student data has to be considered, collected and analysed. As highlighted by Kitchin (2021), "data-driven endeavours are not simply technical systems, but are socio-technical. That is, they are as much a result of human values, desires and social relations as they are scientific principles and technologies" (p. 5).

In the educational context, it is not always clear what data are and how to best make sense of them, i.e., to translate understanding into action. This can be explained by several reasons. One of them relates to the fact that at many occasions, we need to integrate different types of data in order to achieve the intended improvement goals. In an education setting, a diversity of educational tools are frequently used in parallel. One digital system may be used for practicing and taking exams, another - for supporting students' collaborative learning activities. There may also be a variety of learning management systems (LMS) employed for the purpose of delivering an educational module or a course, often including information and instructions, sharing of study materials and collecting student assignments. More advanced

LMS also provide student activity/interaction data which can be useful for understanding reasons behind performance. There are also systems that collect various types of student demographic data which needs to be carefully accounted for to be able to provide *equitable* and *fair* adaptive learning solutions, based on LA measurements (Baker & Hawn, 2021). Equity and fairness are becoming increasingly important goals in education (Mandinach & Schildkamp, *in press*), and in the design of LA systems (e.g., Holstein & Doroudi, 2019; Hakami & Hernandez-Leo, 2020; Wise et al., 2021). An equitable approach adopts an asset-based view which builds upon student strengths, interests and backgrounds (Datnow & Park, 2018).

In sum, many LA researchers argue that in order to improve conditions for learning and to improve teaching and learner support, "the true potential to offer meaningful insights comes from combining data from across different data sources" (Bakharia et al., 2016, p. 378; Mandinach & Gummer, 2016).

The *integration* of data plays a critical role for scalability of LA (Samuelsen et al., 2019). But how do we understand scale of impact in LA? Scholars argue that there are two key tensions in implementing LA for impact (Knight et al., 2020). The first tension refers to population versus learning gain scale. That is, there is a conflict between the ambition to reach as large populations of students as possible and the ambition to make a decisive impact on learning. The second tension relates to the issues of generalizability and adoption, suggesting that hitherto LA researchers have focused on big classes with more general models, "rather than on supporting educators to develop their own analytical tools, and to adopt tools over multiple sites" (p.4).

Scholars furthermore posit that to integrate data from different sources in LA, it is vital to consider the context of the data. In what settings and for what purpose were data generated? By which technological devices were they collected? Integration of data is beneficial for interoperability – semantic, technical, legal, as well as organizational – and may be used to personalise learning and enable better querying and reporting of data (Samuelsen, Chen, & Wasson, 2021).

During the past decade, a change from a heavy focus on accountability towards more emphasis on *continuous improvement of education* has occurred. The focus is on the process of data use within a particular sociocultural context. Educators need to "tap diverse data sources [e.g., demographics, attendance, motivation, and home circumstances] to contextualise student performance and behavior" (Mandinach & Schildkamp, *in press*, p.2) and to reduce 'unfair' algorithmic educational solutions (Baker & Hawn, 2021). Researchers suggest that in order to address representational and measurement bias, i.e., to increase fairness, we [researchers] need – and we need to help teachers to understand how to – collect "better data -- data that includes sufficient proportions of all groups of interests, and where key variables are not themselves biased. This step is recognized as essential among learning analytics practionaries" (Baker & Hawn, 2021, p.14). As stressed by Holstein et al. (2019), one key step for enhancing *fairness* in the algorithms' use in education would be for researchers to find ways to support practitioners, including teachers, in "collecting and curating" higher-quality data sets.

Such data sources can contribute to offering better explanations and contexts to help educators to better understand and interpret what data means.

While it is important to interpret data in view of teaching and learning goals, 'big data' analysis also offers a new channel to understanding. As stressed by Schildkamp (2019), one of the characteristics of big data is that various kinds of data can be linked to each other and that it is possible to look for patterns in these data sets "without having pre-defined hypotheses. In this way, patterns may be discovered that have never been thought of before, which can lead to new possible applications, purposes and goals of data use" (p.261), as well the relevant theoretical developments. Such discovery may lead to new insights and spur new ideas but require caution before application in teaching and learning practice as correlation patterns not necessarily reflect causes and effects.

Another data-related problem concerns a challenge in terms of our understanding of what type of previously unavailable data we need to collect, combine and analyse in order to bring novel insights into students' learning processes. This requires a thorough understanding of the nature of educators' teaching practices and specific teaching-associated problems that the teacher might need help with.

## Conclusions

The review presented here can be summarized by a set of recommendations to learning analytics practice, discussed in various ways in the papers reviewed. Following the Human-Centered Learning Analytics approach, inspired by the user-centered approach since decades taken in the field of human-computer interaction, learning analytics must focus more on understanding of users and use contexts.

1. LA development must start with teaching and learning problems and goals. Do not start with data.
2. Identify data needs based on goals and practice. Do not rest with what is easily available.
3. In order to make sense of data for improving teaching and learning processes, it is important to engage both data professionals and teaching professionals. Inspiration to improvement can come both from data and from practice, but the decisions in learning processes that are to be supported rest with practitioners.
4. Including different data sources and different kinds of data may be necessary so as to capture not only performance but also students, and learning contexts and processes, which are all diverse. However, as this includes both complex situations and qualitative data it is important to tread cautiously and make sure the data analysis used within LA systems is triangulated with empirical studies of use.

5. Involve students and analyze their role in using the data.

Moreover, to be able to scale up LA research and support teachers in their data-driven decision making processes, there is a critical need to develop a *responsible* approach to student data use in education. Even though the LA research community has been for some time interested in the ethics of data-driven practices (e.g., Ferguson, 2019; Tsai et al., 2019), most of this work has been performed in conceptual terms (Arnold & Sclater, 2017). As highlighted by Cerratto-Pargman et al. (2021), research on *applied ethics* has not become pervasive in LA practice. Scholars suggest the so-called 'socially sympathetic' design approach to LA systems (Selwyn, 2019). This approach – is sometimes referred to as 'user-respectful' design as opposed to 'user-centered' only – implies designing LA systems and applications in ways that consider different social contexts in which they are intended to be used, and the different needs and rights of the users. It also requires ensuring informed engagement concerning issues related to privacy, security, and user rights of individuals interacting with these systems (Selwyn, 2019).